\documentclass[pra,twocolumn,showpacs,letterpaper,showpacs,superscriptaddress]{revtex4}
\usepackage{graphicx,amsmath,amssymb,amsfonts,latexsym,color,dcolumn,bm}
\usepackage{appendix}

\begin{document}
\newcommand{\bd}{{\bf d}}      
\newcommand{\bv}{{\bf v}}
\newcommand{\hbp}{\hat{\bp}}
\newcommand{\hbx}{\hat{\bx}}
\newcommand{\hq}{\hat{q}}
\newcommand{\hp}{\hat{p}}
\newcommand{\ha}{\hat{a}}
\newcommand{\had}{{a}^{\dag}}
\newcommand{\ad}{a^{\dag}}
\newcommand{\hsig}{{\hat{\sigma}}}
\newcommand{\nt}{\tilde{n}}
\newcommand{\itf}{\sl}
\newcommand{\eps}{\epsilon}
\newcommand{\bsig}{\pmb{$\sigma$}}
\newcommand{\beps}{\pmb{$\eps$}}
\newcommand{\bmu}{\pmb{$ u$}}
\newcommand{\balpha}{\pmb{$\alpha$}}
\newcommand{\bbeta}{\pmb{$\beta$}}
\newcommand{\bgamma}{\pmb{$\gamma$}}
\newcommand{\bu}{{\bf u}}
\newcommand{\bpi}{\pmb{$\pi$}}
\newcommand{\bSig}{\pmb{$\Sigma$}}
\newcommand{\be}{\begin{equation}}
\newcommand{\ee}{\end{equation}}
\newcommand{\bea}{\begin{eqnarray}}
\newcommand{\eea}{\end{eqnarray}}
\newcommand{\sss}{_{{\bf k}\lambda}}
\newcommand{\ssss}{_{{\bf k}\lambda,s}}
\newcommand{\dip}{\langle\sigma(t)\rangle}
\newcommand{\dipp}{\langle\sigma^{\dag}(t)\rangle}
\newcommand{\sig}{{\tilde{\sigma}}}
\newcommand{\sigd}{{\sigma}^{\dag}}
\newcommand{\sigz}{{\sigma_z}}
\newcommand{\ra}{\rangle}
\newcommand{\la}{\langle}
\newcommand{\om}{\omega}
\newcommand{\Om}{\Omega}
\newcommand{\pa}{\partial}
\newcommand{\bR}{{\bf R}}
\newcommand{\bx}{{\bf x}}
\newcommand{\br}{{\bf r}}
\newcommand{\bE}{{\bf E}}
\newcommand{\bH}{{\bf H}}
\newcommand{\bB}{{\bf B}}
\newcommand{\bP}{{\bf P}}
\newcommand{\bD}{{\bf D}}
\newcommand{\bA}{{\bf A}}
\newcommand{\bek}{{\bf e}\rmk}
\newcommand{\rmk}{_{{\bf k}\lambda}}
\newcommand{\bsij}{{\bf s}_{ij}}
\newcommand{\bk}{{\bf k}}
\newcommand{\bp}{{\bf p}}
\newcommand{\epso}{{1\over 4\pi\eps_0}}
\newcommand{\BB}{{\mathcal B}}
\newcommand{\AAA}{{\mathcal A}}
\newcommand{\NN}{{\mathcal N}}
\newcommand{\mm}{{\mathcal M}}
\newcommand{\RR}{{\mathcal R}}
\newcommand{\bS}{{\bf S}}
\newcommand{\bL}{{\bf L}}
\newcommand{\bJ}{{\bf J}}
\newcommand{\bI}{{\bf I}}
\newcommand{\bF}{{\bf F}}
\newcommand{\bsub}{\begin{subequations}}
\newcommand{\esub}{\end{subequations}}
\newcommand{\baline}{\begin{eqalignno}}
\newcommand{\ealine}{\end{eqalignno}}
\newcommand{\isat}{{I_{\rm sat}}}
\newcommand{\Is}{I^{\rm sat}}
\newcommand{\Ip}{I^{(+)}}
\newcommand{\Imm}{I^{(-)}}
\newcommand{\Inu}{I_{\nu}}
\newcommand{\bInu}{\overline{I}_{\nu}}
\newcommand{\bN}{\overline{N}}
\newcommand{\qnu}{q_{\nu}}
\newcommand{\oqn}{\overline{q}_{\nu}}
\newcommand{\qsat}{q^{\rm sat}}
\newcommand{\Iout}{I_{\nu}^{\rm out}}
\newcommand{\topt}{t_{\rm opt}}
\newcommand{\crr}{{\mathcal{R}}}
\newcommand{\cE}{{\mathcal{E}}}
\newcommand{\cH}{{\mathcal{H}}}
\newcommand{\epsoo}{\epsilon_0}
\newcommand{\ombar}{\overline{\om}}
\newcommand{\cEp}{{\mathcal{E}}^{(+)}}
\newcommand{\cEm}{{\mathcal{E}}^{(-)}}
\newcommand{\bvv}{\tilde{\bv}}
\newcommand{\pr}{^{\prime}}
\newcommand{\dpr}{^{\prime\prime}}
\newcommand{\hk}{\hat{\bk}}
\newcommand{\hn}{\hat{\bf n}}
\title{Simplified derivation of the Kompaneets equation}
\author{Peter W. Milonni}
\affiliation{Theoretical Division, Los Alamos National Laboratory, Los Alamos, New Mexico 87545 USA} 
\affiliation{Department of Physics and Astronomy, University of Rochester, Rochester, NY 14627 USA}
\email{peter_milonni@comcast.net} 
\date{August 26, 2021}
\begin{abstract} An isotropic electromagnetic field in a plasma of thermalized electrons undergoes changes in energy as a result of Compton scattering and an Einstein--Hopf drag force on the electrons, eventually approaching a Bose--Einstein photon distribution at the electron temperature.  The rate of change of field energy due to the combined effects of Compton scattering and the drag force is shown to be described by the Kompaneets equation for photon diffusion in frequency space.  A similarity is noted between this approach and Einstein's derivation of the Planck spectrum based on the recoil of atoms as they absorb and emit radiation.
\end{abstract}
\maketitle

The Kompaneets equation describes the thermal equilibration of radiation via Compton scattering \cite{komp, zeld, olive} in a rarefied plasma in which electrons are thermalized at a temperature $T$.  Under assumptions summarized below, this equation for the average number $n(\om,t)$ of photons at frequency $\om$ at time $t$ is
\be
\frac{\pa n(\om,t)}{\pa t}=\frac{n_e\hbar\sigma_T}{mc}\frac{1}{\om^2}\frac{\pa}{\pa\om}\om^4\Big[n^2(\om)+n(\om)+\frac{k_BT}{\hbar}\frac{\pa n}{\pa\om}\Big],
\label{kompeq}
\ee
where $m$ and $n_e$ are the electron mass and number density and $\sigma_T=(8\pi/3)(e^2/mc^2)^2$ is the cross section for Thomson scattering. Among other applications, the Kompaneets equation has been the basis for analyses of the Sunyaev-Zeldovich effect in the observed cosmic microwave background \cite{sunzel,thorne}. Its derivation is "non-trivial since it has to take account of the interchange of energy between the photons and electrons and also include induced
effects which become important when the occupation number $n$ is large " \cite{longair}. In view of its long history, it is hardly surprising that it has been derived by different methods, typically based on a Fokker--Planck approximation to a Boltzmann or master equation \cite{olive}. The term proportional to $n^2(\om)+n(\om)$ is associated with Compton scattering, including induced scattering, whereas the term proportional to $\pa n/\pa\om$ describes a photon diffusion in frequency space and is attributable to ``the heating of the photons by hot electrons" \cite{thorne}. 

The purpose of this note is to present a simpler derivation and interpretation of the Kompaneets equation based on the rate of change of field energy due to the combined effects of Compton scattering and a drag force the field exerts on the electrons. In particular, it is shown that the term proportional to
$\pa n/\pa\om$ appears because of an Einstein--Hopf drag force acting on an electron as it moves in an isotropic radiation field. The term proportional to $n^2(\om)+n^2(\om)$ accounts for the momentum fluctuations of an electron due to Compton scattering in such a field in which the photon number variance is given by the Bose--Einstein formula, $\la\Delta n^2(\om)\ra=n^2(\om)+n(\om)$. We assume an isotropic, unpolarized field with a spectral energy density $\rho(\om)$ and, as in Kompaneets's original work and in most of the work it has inspired, treat the problem nonrelativistically. The latter approximation allows us to derive the drag force based on the nonrelativistic form of the radiation reaction field acting on an electron, and to use the Thomson cross section for photon scattering. The photons are assumed to undergo only small relative changes in frequency in each collision with an electron. We assume a rarefied, non-interacting plasma in which the electrons are in thermal equilibrium at temperature $T$ and remain so at all times \cite{lowell}. It is assumed that absorption and emission are negligible, so that photon number is conserved.

We first derive a formula for the drag force on an electron moving along a single direction $x$ in a single-mode electric field $E_z$ having a frequency $\om$ and pointing in a direction $z$ orthogonal to $x$.  The field exerts on the electron a force 
\be
F_x=\frac{e}{2}z\frac{\pa E_z}{\pa x}+ \frac{e}{2}\frac{\pa E_z}{\pa x}z
\label{eqq}
\ee 
along the $x$ direction, where $z$ is the field-induced displacement of the electron along the $z$ direction. In the quantum-mechanical approach taken here, $z$ and $E_z$ are Hermitian operators, and the electric field operator at the electron's position  $\br$ is
\be
E_z=i\big(\frac{2\pi\hbar\om}{V}\big)^{1/2}\big[ae^{-i(\om t-\bk\cdot\br)}-\had e^{i(\om t-\bk\cdot\br)}\big]
\label{eq2}
\ee
in standard notation in which $a$ and $\had$ are photon annihilation and creation operators, respectively, and $V$ is a quantization volume. With this field,
\be
ez=i\big(\frac{2\pi\hbar\om}{V}\big)^{1/2}\big[\alpha(\om)ae^{-i(\om t-\bk\cdot\br)}-\alpha^*(\om)\had e^{i(\om t-\bk\cdot\br)}\big],
\label{eq3}
\ee
where $\alpha(\om)$ is the (complex) ``polarizability." Therefore,
\be
F_x=4\pi k_x\alpha_I(\om)\Big[\frac{\hbar\om}{V}(\had a+\frac{1}{2})\Big],
\label{eq4}
\ee
where $\alpha_I(\om)$ is the imaginary part of $\alpha(\om)$. We have dropped terms
$aa$ and $\had\had$, which, as is easily shown, make no contribution to the force in a thermal field. The 1/2 appears as a consequence of the commutation relation $[a,\had]=1$ and the zero-point ($T=0$) field, but, as shown below, this field likewise exerts no average force on the particle. Therefore we replace (\ref{eq4}) by
\be
F_x=4\pi k_x\alpha_I(\om)\frac{\hbar\om}{V}\had a.
\label{eq5}
\ee
The average of $F_x$ over the state of the field is
\be
F_x=4\pi k_x\alpha_I(\om)\frac{\hbar\om}{V}\la\had a\ra = 4\pi k_x\alpha_I(\om)\Big[\frac{\hbar\om}{V}n(\om)\Big],
\label{eq6}
\ee
where $n(\om)$ is the average number of photons in the single-mode field. The term in brackets is the field energy density. We now generalize to 
 radiation with energy density $\rho(\om)d\om$ in the frequency interval $[\om,\om+d\om]$ and  propagating at the angle 
$\theta$ with respect to the $x$ axis ($k_x=(\om/c)\cos\theta$):
\be
F_x=\frac{4\pi}{c} \om\alpha_I(\om)\rho(\om)\cos\theta\frac{d\Om}{4\pi} d\om,
\label{eq7}
\ee
where $d\Om$ is the solid angle subtended by this radiation \cite{polarization}. Next we allow for all field frequencies and propagation directions:
\be
 F_x=\frac{4\pi}{c}\int_0^{\infty}d\om\om\alpha_I(\om)\rho(\om)\frac{1}{4\pi}\int_0^{2\pi}d\phi\int_0^{\pi}d\theta\sin\theta\cos\theta.
\label{eq8}
\ee
$F_x=0$, since we are assuming an isotropic field in which $\rho$ is independent of propagation direction.  

Now we allow the particle to move with velocity $v$ along the $x$ direction. Then it does not see an isotropic spectral
energy density $\rho(\om)$ of the field, but rather a spectral energy density given by \cite{ford,pwmoup} 
\be
\rho^{\prime}(\om^{\prime},\theta^{\prime})\cong \big(1-\frac{3v}{c}\cos\theta^{\prime}\big)\Big[\rho(\om^{\prime})+\frac{v}{c}\om^{\prime}\frac{d\rho(\om^{\prime})}{d\om^{\prime}}\cos\theta^{\prime}\Big]
\label{eq9}
\ee
to lowest order in $v/c$. Here $\om^{\prime}$ is the Doppler-shifted frequency and $\theta^{\prime}$ is the (aberrated) angle seen by the moving particle
($\cos\theta^{\prime}\cong\cos\theta-\frac{v}{c}\sin^2\theta$). Thus, for the moving particle,
\be
F_x=m\frac{dv}{dt}=-\xi v,
\label{eq10a}
\ee
\be
\xi\cong\frac{1}{c}\int_0^{\infty}d\om\om\alpha_I(\om)\int_0^{2\pi}d\phi\int_0^{\pi}d\theta^{\prime}\sin\theta^{\prime}\cos\theta^{\prime}\rho(\om,\theta^{\prime}).
\ee
Doing the integrations, we obtain the drag coefficient 
\be
\xi=\frac{4\pi}{c^2}\int_0^{\infty}d\om\om\alpha_I(\om)\big[\rho(\om)-\frac{\om}{3}\frac{d\rho}{d\om}\big]
\label{eq10}
\ee
to lowest order in $v/c$. This simply generalizes the expression originally derived by Einstein and Hopf \cite{einhopf}, who treated the particle as a linear oscillator with a polarizability characterized by a single resonance frequency \cite{mkr}. 

From the classical Abraham-Lorentz equation of motion 
\be
m(\ddot{\br}-\tau_e\stackrel{...}{\br})=\bE
\ee
for an electron in an electric field, including the radiation reaction field of the electron, 
\be
\alpha(\om)=-\frac{e^2/m}{\om^2+i\tau_e\om^3},
\ee
\be
\alpha_I(\om)=\frac{e^2\tau_e\om^3/m}{\om^4+\tau_e^2\om^6}.
\label{eq12a}
\ee
Since $\tau_e=2e^2/3mc^3\cong 6.3\times 10^{-24}$ s, we approximate (\ref{eq12a}) by 
\be
\alpha_I(\om)=\frac{e^2\tau_e}{m\om} = \frac{c\sigma_T}{4\pi\om},
\label{eq12b}
\ee
where again $\sigma_T$ is the cross section for Thomson scattering. Then 
\be
\xi=\frac{\sigma_T}{c}\int_0^{\infty}d\om\big[\rho(\om)-\frac{\om}{3}\frac{\pa\rho}{\pa\om}\big].
\ee
Finally we  write the spectral energy density in terms of the number of modes $ \om^2/\pi^2c^3$ per unit volume and frequency interval, the photon energy $\hbar\om$, and the number of photons $n(\om)$ at frequency $\om$:
\be
\rho(\om)=(\hbar\om^3/\pi^2c^3)n(\om),
\label{rhoeq}
\ee
\be
\rho(\om)-\frac{\om}{3}\frac{\pa\rho}{\pa\om}=-\frac{\hbar}{3\pi^2c^3}\om^4\frac{\pa n}{\pa\om},
\label{rela}
\ee
and 
\be
\xi=-\frac{\hbar\sigma_T}{3\pi^2c^4}\int_0^{\infty}d\om\om^4\frac{\pa n}{\pa\om}.
\label{eqxi}
\ee
We note that, for $T=0$, $n(\om)=1/2$ and therefore $\xi=0$.

For the isotropic field we are assuming, equation (\ref{eqxi}) gives the drag force ${\bf F}_D=-\xi{\bf v}$ on an electron moving with the velocity ${\bf v}$ in any direction. In the case of $n_e$ electrons per unit volume, this force results in a rate of loss of kinetic energy per unit volume 
$n_e\la{\bf F}_D\cdot{\bf v}\ra$ and a rate of increase of field energy density given by
\bea
\Big\la\frac{dE}{dt}\Big\ra_{\rm D}&=&-n_e\la{\bf F}_D\cdot{\bf v}\ra=n_e\big\la\xi {\bf v}^2\ra=3n_e\frac{k_BT}{m}\xi\nonumber\\
&&\mbox{}=-\frac{n_e\hbar k_BT\sigma_T}{\pi^2mc^4}\int_0^{\infty}d\om\om^4\frac{\pa n}{\pa\om},
\eea
where we have used $\la\frac{1}{2}m{\bf v}^2\ra=\frac{3}{2}k_BT$ to express this rate in terms of the electron temperature $T$. 

Consider now the effect of Compton scattering on the electrons' momentum fluctuations. An incoming photon of momentum $\hbar\bk=k\hat{\bk}=(\hbar\om/c)\hat{\bk}$ and scattered by an electron results in an outgoing photon of momentum $\hbar\bk^{\prime}=(\hbar\om/c)\hat{\bk}^{\prime}$ in the electron's rest frame, where $\hat{\bk}$  and $\hat{\bk}^{\prime}$ are unit vectors. The electron experiences a change in its momentum squared,
$ \Delta P^2=(\hbar\om/c)^2(\bk-\bk^{\prime})^2$. Let $N(\om)d\om d\Om$ be the number of incident photons per unit area and per unit time in the frequency and solid angle  ranges $[\om,\om+d\om]$ and $[\Om,\Om+d\Om]$, respectively. If $\rho(\om)$ is the spectral energy density of the isotropic field, such that $\rho(\om)d\om d\Om/4\pi$ is the field energy per unit volume in the frequency and solid angle ranges $[\om,\om+d\om]$ and $[\Om,\Om+d\Om]$, $N(\om)=c\rho(\om)/4\pi\hbar\om$. Let $|f(\bk,\bk^{\prime})|^2$ be the differential cross section for scattering in which a photon's momentum is changed from $\bk$ to $\bk^{\prime}$, $|\bk|=|\bk^{\prime}|=k$. Then the change in the mean-square momentum $\la \Delta P^2\ra$ in a time $\Delta t$ is obtained by integrating over all possible frequencies and  propagation directions of the incoming photon and over all possible propagation directions of a single outgoing photon:
\bea
\frac{\la\Delta P^2\ra}{\Delta t}&=&\int_0^{\infty} d\om\int d\Om\frac{c\rho(\om)}{4\pi\hbar\om}\int d\Om^{\prime}|f(k\hat{\bk},k\hat{\bk})|^2\nonumber\\
&&\mbox{}\times\big(\frac{\hbar\om}{c}\big)^2[\hat{\bk}^{\prime}-\hat{\bk}]^2,
\eea
or
\bea
\frac{\la\Delta P^2\ra}{\Delta t}&=&\frac{\hbar}{c}\int_0^{\infty}d\om\om\rho(\om)\int_0^{\pi}d\theta\sin{\theta}|f(\bk,\bk^{\prime})|^2\nonumber\\
&&\mbox{}\times (1-\cos{\theta}),
\label{eq100}
\eea
where $\theta$ is the angle between $\hat{\bk}$ and $\hat{\bk}^{\prime}$.

We can express (\ref{eq100}) in terms of the average photon number $n(\om)$  using equation (\ref{rhoeq}): 
\bea
\frac{\la\Delta P^2\ra}{\Delta t}&=&\frac{\hbar^2}{\pi^2 c^4}\int_0^{\infty}d\om\om^4n(\om)\nonumber\\
&&\mbox{}\times\int_0^{\pi}d\theta\sin{\theta}|f(\bk,\bk^{\prime})|^2(1-\cos{\theta}).
\label{eq101}
\eea
This expression is applicable to scattering in which an average number $n(\om)$ of photons at each frequency of an isotropic, unpolarized field is incident on the particle and a single photon of approximately the same frequency is scattered. But in the case of photons satisfying Bose-Einstein statistics,
such as photons of blackbody radiation, (\ref{eq101}) must be modified to allow for {\sl induced} scattering: if there are $n(\om)$ incoming photons in a mode of frequency $\om$,  the rate of scattering into that same mode is enhanced by the factor $n(\om)+1$ . To allow for both ``spontaneous" scattering
and induced scattering we replace (\ref{eq101}) by
\bea
\frac{\la\Delta P^2\ra}{\Delta t}&=&\frac{\hbar^2}{\pi^2 c^4}\int_0^{\infty}d\om\om^4\big[n(\om)(n(\om)+1)\big]\nonumber\\
&&\mbox{}\times\int_0^{\pi}d\theta\sin{\theta}|f(\bk,\bk^{\prime})|^2(1-\cos{\theta}).
\label{eq102}
\eea

The differential cross section of interest is given by the well-known formula for Rayleigh scattering:
\be
|f(\bk,\bk^{\prime})|^2=\Big(\frac{\om}{c}\Big)^4|\alpha(\om)|^2\frac{1}{2}(1+\cos^2\theta).
\ee
We also use the relation
\be
\alpha_I(\om)=\frac{2}{3}\Big(\frac{\om}{c}\Big)^3|\alpha(\om)|^2
\ee
that follows from the optical theorem for Rayleigh scattering \cite{pwmoup}. Then, carrying out the integration over $\theta$ in (\ref{eq102}), we obtain
\be
\frac{\la\Delta P^2\ra}{\Delta t}=\frac{8\hbar^2}{\pi c^5}\int_0^{\infty}d\om\om^5\alpha_I(\om)\big[n(\om)(n(\om)+1)\big].
\label{eq103}
\ee
We again use  (\ref{eq12b}) for $\alpha_I(\om)$ and obtain
\be
\frac{\la\Delta P^2\ra}{\Delta t}=\frac{2\hbar^2\sigma_T}{\pi^2 c^4}\int_0^{\infty}d\om\om^4\big[n^2(\om)+n(\om)\big].
\label{eqmom}
\ee
The field energy per unit volume due to Compton scattering therefore changes at a rate
\bea
\Big\la\frac{dE}{dt}\Big\ra_{\rm C}&=&-\frac{n_e}{2m}\frac{\la\Delta P^2\ra}{\Delta t}\nonumber\\
&=&-\frac{\hbar^2n_e\sigma_T}{m\pi^2c^4}\int_0^{\infty}d\om\om^4[n^2(\om)+n(\om)].
\eea

The total rate at which the field energy density changes is  
\be
\Big\la\frac{dE}{dt}\Big\ra=\Big\la\frac{dE}{dt}\Big\ra_{\rm D}+\Big\la\frac{dE}{dt}\Big\ra_{\rm C}=-\hbar\int_0^{\infty}d\om\om^4g(\om).
\label{eqgood}
\ee
Here
\be
g(\om)\equiv \frac{n_e\hbar\sigma_T}{mc}\Big[n^2(\om)+n(\om)+\frac{k_BT}{\hbar}\frac{\pa n}{\pa\om}\Big],
\ee
which vanishes when $n(\om)=1/[e^{(\hbar\om-\mu)/k_BT}-1]$, where $\mu$ is the chemical potential. 
Assuming $\om^5g(\om)\rightarrow 0$ as $\om\rightarrow\infty$, we integrate by parts to re-write (\ref{eqgood}) as
\be
\Big\la\frac{dE}{dt}\Big\ra=\hbar\int_0^{\infty}d\om\om\frac{\pa}{\pa\om}(\om^4g).
\ee
Expressing the field energy density as $\int_0^{\infty}d\om(\hbar\om^3/\pi^2c^3)n(\om,t)$, we then have
\be
\int_0^{\infty}d\om\frac{\hbar\om^3}{\pi^2c^3}\frac{\pa n(\om,t)}{\pa t}=\hbar\int_0^{\infty}d\om\om\frac{\pa}{\pa\om}(\om^4g),
\ee
from which we deduce the Kompaneets equation (\ref{kompeq}).

Finally we note that, from the relations (\ref{eq12a})--(\ref{rela}),
\be
\Big\la\frac{dE}{dt}\Big\ra_{\rm D}=\frac{12n_e\pi k_BT}{mc^2}\int_0^{\infty}d\om\om\alpha_I(\om)\big[\rho(\om)-\frac{\om}{3}\frac{\pa\rho}{\pa\om}\big]
\ee
and
\be
\Big\la\frac{dE}{dt}\Big\ra_{\rm C}=-\frac{4n_e\pi^3c}{m}\int_0^{\infty}\frac{\alpha_I(\om)}{\om}\big[\rho^2(\om)+\frac{\hbar\om^3}{\pi^2c^3}\rho(\om)\big].
\ee
In equilibrium these two expressions must add to zero, implying 
\be
\rho(\om)-\frac{\om}{3}\frac{\pa\rho}{\pa\om}=\frac{\pi^2c^3}{3k_BT\om^2}\big[\rho^2(\om)+\frac{\hbar\om^3}{\pi^2c^3}\rho(\om)\big].
\label{eingenius}
\ee
The solution of this differential equation with $\rho(0)=0$ is the Planck spectrum. Equation (\ref{eingenius}) is exactly the equation derived by Einstein in his theory of blackbody radiation based on the drag force and momentum fluctuations---not for free electrons as they scatter radiation, but for two--level {atoms} as they absorb and emit radiation \cite{einstein}. Einstein assumes that the atoms' average kinetic energy is $(3/2)k_BT$, but that the occupation probabilities of the atomic levels change as the atoms absorb and emit radiation. In the simple approach desribed in this note the particles (electrons) have no internal degrees of freedom affected by interaction with the field. The average kinetic energy is fixed at $(3/2)k_BT$ independently of the field, whose approach to thermal equilibrium is then described by the Kompaneets equation.
\\ \\
{Acknowledgement}: I thank Dr. Gennady P. Berman for helpful comments.
\\ \\
{Author Declarations}: The author has no conflict of interest to disclose.
\\ \\
Data Availability Statement: Data sharing is not applicable to this article as no new data were created or analyzed in this study.


\begin{thebibliography}{99}
\bibitem{komp} A. S. Kompaneets, ``The establishment of thermal equilibrium between quanta and electrons," Sov. Phys. JETP {\bf 4}, 730 (1957).
\bibitem{zeld} For a discussion of the Kompaneets equation and related topics, including a history of earlier developments, see Ya. B. Zel'dovich, ``Interaction of free electrons with electromagnetic radiation," Sov. Phys. Usp. {\bf  18}, 79 (1975).
\bibitem{olive} For a more recent review and analysis see G. E. F. Oliveira, C. Maes, and K. Meerts, ``On the derivation of the Kompaneets equation," arXiv:2103.06654 astro-ph (March, 2021) and references therein.
\bibitem{sunzel} R. A. Sunyaev and Ya. B. Zel'dovich,  ``Small-scale fluctuations of relic radiation," Astrophysics and Space Science. {\bf 7}, 3 (1970);  
R. A. Sunyaev and Ya. B. Zel'dovich, ``Microwave background radiation as a probe of the contemporary structure and history of the universe," Ann. Rev. Astron. and Astrophys. {\bf 18}, 537 (1980). 
\bibitem{thorne} See, for instance, K. S. Thorne and R. D. Blandford, {\sl Relativity and Cosmology: Volume 5 of Modern Classical Physics} (Princeton University Press, 2017), Section 28.6.3.
\bibitem{longair} M. S. Longair, {\sl High Energy Astrophysics}, third edition (Cambridge University Press, 2011), p. 250.
\bibitem{lowell} It has been shown by L. S. Brown, Ann. Phys. {\bf 200}, 190 (1990), that the Kompaneets equation applies as well to an interacting plasma, provided that the electrons are in thermal equilibrium.
\bibitem{polarization} We assume that  $\rho(\om)$ describes an unpolarized field, so that a factor of 2 allowing for two field polarizations orthogonal to
each direction of propagation is included in the defnition of $\rho(\om)$.
\bibitem{ford} This formula follows from the Lorentz transformation of the spectral energy density of an isotropic field when the Lorentz factor $\gamma=\sqrt{1-v^2/c^2}\cong 1$. This Lorentz-transformed spectral energy density is given by equation (17) of G. W. Ford and R. F. O'Connell, Phys. Rev. E {\bf 88}, 044101 (1993). 
\bibitem{pwmoup} See also, for instance, P. W. Milonni, {\sl An Introduction to Quantum Optics and Quantum Fluctuations} (Oxford University
Press, 2019), Sections 1.11 and 2.8 and references therein.
\bibitem{einhopf} A. Einstein and L. Hopf, ``Statistical investigation of a resonator's motion in a radiation field," Ann.  Physik {\bf 33}, 1105 (1910). 
\bibitem{mkr} An expression equivalent to (\ref{eq10}) has been derived in a different way by V. Mkrtchian {\sl et al.}, ``Universal thermal radiation drag on neutral objects," Phys. Rev. Lett. {\bf 91}, 220801 (2003). The authors were evidently unaware of the work of Einstein and Hopf.
\bibitem{einstein} A. Einstein, "Quantum theory of radiation," Phys. Zs. {\bf 18}, 121 (1917).
\end{thebibliography}
\end{document}